\documentclass[aps,floats]{revtex4-2}
\usepackage{amsmath,amssymb}
\usepackage{graphicx,epsfig}
\usepackage[greek,english]{babel}
\usepackage{bbold}

\begin{document}
\bibliographystyle {plain}

\pdfoutput=1
\def\oppropto{\mathop{\propto}} 
\def\opsimeq{\mathop{\simeq}}
\def\opoverderline{\mathop{\overline}}
\def\operarrow{\mathop{\longrightarrow}}
\def\opsim{\mathop{\sim}}

\def\opmin{\mathop{\min}} 
\def\opmax{\mathop{\max}} 
\def\oplim{\mathop{\lim}}

\title{ Markov dualities via the spectral decompositions of the two Markov generators\\
in their bi-orthogonal basis of right and left eigenvectors } 


\author{C\'ecile Monthus}
\affiliation{Universit\'e Paris-Saclay, CNRS, CEA, Institut de Physique Th\'eorique, 91191 Gif-sur-Yvette, France}


\begin{abstract}
The notion of Markov duality between two Markov processes that can live in two different configurations spaces $(x,{\tilde x})$ is revisited via the spectral decompositions of the two Markov generators in their bi-orthogonal basis of right and left eigenvectors. In this formulation, the two generators should have the same eigenvalues $(-E)$ that may be complex, while the duality function $\Omega(x,{\tilde x})$ can be considered as a mapping between the right and the left eigenvectors of the two models. We describe how this spectral perspective is useful to better understand two well-known dualities between processes defined in the same configuration space: the Time-Reversal duality corresponds to an exchange between the right and the left eigenvectors that involves the steady state, while in the Siegmund duality, the left eigenvectors correspond to integrals of the dual right eigenvectors. We then focus on the famous Moment-Duality between the Wright-Fisher diffusion on the interval $x \in [0,1] $ and the Kingman Markov jump process on the semi-infinite lattice $n \in {\mathbb N}$ in order to analyze the relations between their eigenvectors living in two different configuration spaces. Finally, we discuss how the spectral perspective can be used to construct new dualities and we give an example for the case of non-degenerate real eigenvalues, where one can always construct a dual Directed Jump process on the semi-infinite lattice $n \in {\mathbb N}$, whose transitions rates are the opposite-eigenvalues.

\end{abstract}

\maketitle


\section { Introduction}

The notion of Markov dualities
(see the reviews \cite{ReviewMohle,ReviewDuality,AlgebraicReview} with various scopes and references therein)
has attracted a lot of interest recently in many types of models
\cite{tailleurMapping,giardina_particle,giardina_transport,Redig_genetic,DualityEigen,DualityHidden},
while
the oldest example is known under the name of Siegmund duality 
\cite{siegmund,Cox1983,cliff,Dette,siegmund_Intertwining,Kolo,siegmund_pathwise,Lorek,Zhao,{TheoExchangeSiegmund}} with the recent generalization to several active models \cite{siegmund_runtumble,siegmund_bridge},
but has also been analyzed within various other perspectives 
\cite{levy,Ciesielski-Taylor,biane,toth,tourigny}.

The most standard definition of Markov duality (see the reviews \cite{ReviewMohle,ReviewDuality,AlgebraicReview} and references therein)
is that the process $X(t)$ starting at $X(t=0)=x_0$ and the process ${\tilde X}(t)$ starting at ${\tilde X}(t=0)={\tilde x_0}$
are dual with respect to the duality function $\Omega(x,\tilde x)$ that involves the coordinates $(x,\tilde x) $ of the two configurations spaces
if the two following averaged values coincide for any time $t$ and for any initial conditions $x_0$ and ${\tilde x_0}$
\begin{eqnarray}
 {\mathbb E}_{x_0} [\Omega(X(t), {\tilde x}_0) ]   = {\mathbb E} _{\tilde x_0} [\Omega(x_0, {\tilde X}(t))  ]
\label{defdualityfonction}
\end{eqnarray}
This formulation has the advantage of producing concrete results for the observables $\Omega(x,\tilde x) $,
but has the drawback of giving the impression that "finding dual processes is something of a black art"
as quoted in the introduction of the review \cite{ReviewDuality}.
As a consequence, various reformulations of the notion of Markov duality have been studied
(see the reviews \cite{ReviewMohle,ReviewDuality,AlgebraicReview} and references therein).
For instance in the algebraic approach for many-particle models, 
the two Markov generators are rewritten in terms of elementary operators
 that satisfy the same commutations relations \cite{ReviewDuality,AlgebraicReview,tailleurMapping,giardina_particle,giardina_transport,Redig_genetic,DualityEigen,DualityHidden}. Some other Markov dualities can also be re-interpreted via the supersymmetric properties of
Markov generators \cite{c_boundarydriven,c_susySVD,c_susyFokkerPlanck,mathSusyScaleSpeed}.
 
 In the present paper, we will focus on the spectral approach that can be applied to any Markov duality as follows.
 The definition of Eq. \ref{defdualityfonction} can be first translated into the standard operator identity 
 that will be recalled in Eq. \ref{av12}
 and that only involves the duality function $\Omega(.,.)$
  and the two Markov generators, while the time $t$ and the two initial conditions $x_0$ and ${\tilde x}_0$
  have disappeared (see the reviews \cite{ReviewMohle,ReviewDuality,AlgebraicReview} and references therein
  where this operator identity is written in various forms depending on the notations).
  This operator identity 
  can be then further analyzed in terms of the spectral decompositions of the two
  Markov generators, as explained in the mathematical paper \cite{DualityEigen}
  containing propositions, theorems and proofs, in particular for the most general case involving Jordan blocks.
  The goal of the present paper is to revisit this spectral approach in the more informal style of physicists
  using the standard bra-ket notations familiar from quantum mechanics. 
  To be more concrete and to obtain simpler results, we will only consider 
  the cases of Markov jump processes and diffusion processes 
  whose generators have non-degenerate discrete eigenvalues and can be diagonalized
  in the bi-orthogonal basis of their right and left eigenvectors. 
   In addition, we will discuss the application to various examples
     that can be considered as complementary with respect to the examples analyzed in the mathematical paper \cite{DualityEigen}.


The paper is organized as follows.
In section \ref{sec_spectral}, we describe how the notion of Markov duality of Eq. \ref{defdualityfonction}
can be reformulated in terms the spectral decompositions of the two Markov generators
 in the bi-orthogonal basis of their right and left eigenvectors.
In section \ref{sec_TimeSiegmund}, two well-known Markov dualities between processes defined in the same configuration space, namely the Time-Reversal-Duality and the Siegmund-Duality,
are revisited via the spectral perspective in order to write the simple links between the right and left eigenvectors
of the two generators.
In section \ref{sec_moment}, the Moment-Duality between the Wright-Fisher diffusion on the interval $x \in [0,1] $ and the Kingman Markov jump process on the semi-infinite lattice $n \in {\mathbb N}$ is analyzed from the spectral point of view, and various generalizations are studied.
In section \ref{sec_DirectedEigen}, we discuss how the spectral form of the duality function 
can be used to construct new dualities and we give a simple example for the case of non-degenerate real eigenvalues,
where one can always construct a dual Directed Jump process on the semi-infinite lattice $n \in {\mathbb N}$.
Our conclusions are summarized in section \ref{sec_conclusion}, while Appendix \ref{app_Directed}
contains the detailed analysis of the Directed Markov jump process on the semi-infinite lattice $n \in {\mathbb N}$
with arbitrary transitions rates, whose spectral properties are used in two sections of the main text.


\section{ Spectral analysis of the Markov duality between two generators  }

\label{sec_spectral}

In this section, we describe why it is useful to analyze the notion of Markov duality 
between two Markov processes via the spectral decompositions of the two Markov generators
 in the bi-orthogonal basis of their right and left eigenvectors.

\subsection{ Notations for the generators and for the propagators of the two Markov processes    }

It is convenient to use the bra-ket notations of quantum mechanics as follows :

(i) the orthonormalized basis $\vert x \rangle$ denotes the configuration space of the first Markov process
of generator ${\bold w} $
 in order to write the propagator $p_t(x \vert x_0) $ as
\begin{eqnarray}
p_t(x \vert x_0)  = \langle x \vert e^{t {\bold w} } \vert x_0 \rangle
\label{propagator}
\end{eqnarray}

(ii) the orthonormalized basis $\vert {\tilde x} \rangle \! \rangle$ involving double-bra-ket
denotes the configuration space of the second Markov process  
of generator ${\tilde{\bold w}} $
 in order to write its propagator ${\tilde p}_t({\tilde x} \vert  {\tilde x}_0) $ as
\begin{eqnarray}
{\tilde p}_t({\tilde x} \vert  {\tilde x}_0) \equiv 
 \langle  \! \langle {\tilde x} \vert e^{t {\tilde{\bold w}}} \vert {\tilde x}_0 \rangle \! \rangle
\label{propagatortilde}
\end{eqnarray}

In this paper, we will consider two types of Markov generators that are described in the two following subsections.


\subsubsection{ Markov matrix $w(x,y)=\langle x \vert {\bold w} \vert y \rangle $ for Markov jump processes in a discrete configuration space }

When the configuration space is discrete, the orthonormalization and the closure
of the configuration basis $\vert x \rangle$ reads
\begin{eqnarray}
 \langle  x \vert  y \rangle && = \delta_{x,y}
\nonumber \\
\sum_x \vert  x \rangle \langle x \vert && = {\bold 1} 
\label{orthonormaxdiscrete}
\end{eqnarray}

 The master equation for the probability $p_t(x\vert x_0) $ to be at configuration $x$ at time $t$
 when starting at $x_0$ at time $t=0$
\begin{eqnarray}
 \partial_t p_t(x\vert x_0) && = \sum_{y } w(x,y) p_t( y \vert x_0) = w(x,x)  p_t( x\vert x_0 ) + \sum_{y \ne x} w(x,y) p_t( y \vert x_0)
\label{master}
\end{eqnarray}
involves the matrix elements $w(x,y) $ of the Markov matrix $ {\bold w}$
\begin{eqnarray}
w(x,y)=\langle x \vert {\bold w} \vert y \rangle
\label{wboldjump}
\end{eqnarray}
with the following interpretation :

(i) The off-diagonal element $w(x,y) \geq 0 $ represents the transition rate
from $y$ towards another configuration $x \ne y$

(ii) The diagonal element $w(x,x) $ is negative and determined by the off-diagonal elements
\begin{eqnarray}
w(x,x)  = - \sum_{y \ne x} w(y,x) <0
\label{wdiag}
\end{eqnarray}


\subsubsection{ Fokker-Planck differential generator for diffusion processes on the interval $x \in [0,L] $}

For diffusion processes on the one-dimensional interval $x \in [0,L]$,
 the orthonormalization and the closure of Eq. \ref{orthonormaxdiscrete}
becomes for the continuous configuration basis $\vert  x \rangle$ 
\begin{eqnarray}
 \langle  x \vert  y \rangle && = \delta ( x- y)
 \nonumber \\
 \int_{0}^{L} d x \vert  x \rangle \langle  x \vert   && = {\bold 1} 
\label{orthonormaxcontinuous}
\end{eqnarray}

The master Equation \ref{master} is replaced by the Fokker-Planck dynamics for the probability 
$p_t(x \vert x_0)$
that involves the Ito force $F_I(x)$ and the diffusion coefficient $D(x)$
\begin{eqnarray}
 \partial_t p_t(x\vert x_0) && = - \frac{\partial}{\partial x } \bigg( F_I(x) p_t(x \vert x_0) \bigg) + 
 \frac{\partial^2}{\partial x^2 } \bigg( D(x) p_t(x \vert x_0) \bigg) 
 \equiv {\cal F} p_t(x \vert x_0)
\label{FokkerPlanck}
\end{eqnarray}
So the Fokker-Planck differential operator $ {\cal F}$ replaces the Markov matrix ${\bold w}$,
while the adjoint matrix ${\bold w}^{\dagger}$ is replaced by
the adjoint operator
\begin{eqnarray}
{\cal F}^{\dagger}  = F_I(x)  \frac{\partial}{\partial x } + D(x)  \frac{\partial^2}{\partial x^2 } 
\label{FokkerPlanckAdjoint}
\end{eqnarray}


\subsubsection{ Notations for the remainder of the present section }

In the remainder of the present section concerning the general analysis of Markov dualities, 
we will consider that the duality is between two Markov jump processes in discrete spaces
governed by the Markov matrices ${\bold w}$ and ${\tilde{\bold w}} $, but everything can be translated to 
the case of diffusion processes in continuous spaces with Fokker-Planck generators, as will be described in the examples considered in further sections.


\subsection{ Markov duality as an operator identity involving the two Markov generators  }


\subsubsection{ Reformulation of the duality of Eq. \ref{defdualityfonction} as an operator identity 
involving the two Markov generators}

As recalled in the Introduction, the notion of duality is based on a duality function $\Omega(x,{\tilde x}) $
 that involves the two configurations spaces $x$ and ${\tilde x} $,
so that it is convenient to consider 
that it is a matrix element between the bra $\langle x \vert $ and the ket $\vert {\tilde x} \rangle \! \rangle $
of the matrix ${\bold \Omega}$
\begin{eqnarray}
\Omega(x,{\tilde x}) \equiv \langle x \vert {\bold \Omega} \vert {\tilde x} \rangle \! \rangle 
\label{dualityfunction}
\end{eqnarray}

Then the two averaged values that appear in the definition of Eq. \ref{defdualityfonction}
can be rewritten in terms of the propagators of the two processes as follows:
 
(i) the averaged value on the left handside of Eq. \ref{defdualityfonction}
 can be computed with the propagator $p_t(x \vert x_0) = \langle x \vert e^{t {\bold w} } \vert x_0 \rangle$ 
 of Eq. \ref{propagator} for the first Markov process
\begin{eqnarray}
{\mathbb E}_{x_0} [\Omega(X(t), {\tilde x}_0) ]  &&  
=  \sum_x \langle x \vert {\bold \Omega} \vert {\tilde x}_0 \rangle \! \rangle p_t(x \vert x_0)
= \sum_x 
  \langle x \vert {\bold \Omega} \vert {\tilde x}_0 \rangle \! \rangle \langle x \vert e^{t {\bold w} } \vert x_0 \rangle
=\sum_x 
 \langle x_0 \vert e^{t {\bold w}^{\dagger} } \vert x \rangle \langle x \vert {\bold \Omega} \vert {\tilde x}_0 \rangle \! \rangle
\nonumber \\
&& =  \langle x_0 \vert e^{t {\bold w}^{\dagger} }  {\bold \Omega} \vert {\tilde x}_0 \rangle \! \rangle
\label{av1}
\end{eqnarray}

(ii) the averaged value on the right handside of Eq. \ref{defdualityfonction}
can be computed with the propagator ${\tilde p}_t({\tilde x} \vert  {\tilde x}_0) =\langle  \! \langle {\tilde x} \vert e^{t {\tilde{\bold w}}} \vert {\tilde x}_0 \rangle \! \rangle $
of Eq. \ref{propagatortilde} for the second Markov process
\begin{eqnarray}
 {\mathbb E} _{\tilde x_0} [\Omega(x_0, {\tilde X}(t))  ] 
 && = \sum_{\tilde x}  \langle x_0 \vert {\bold \Omega} \vert {\tilde x} \rangle \! \rangle  
 {\tilde p}_t({\tilde x} \vert  {\tilde x}_0)
 = \sum_{\tilde x}  \langle x_0 \vert {\bold \Omega} \vert {\tilde x} \rangle \! \rangle  
\langle  \! \langle {\tilde x} \vert e^{t {\tilde{\bold w}}} \vert {\tilde x}_0 \rangle \! \rangle
\nonumber \\
&&=  \langle x_0 \vert {\bold \Omega} e^{t  {\tilde{\bold w}} } \vert {\tilde x}_0 \rangle \! \rangle 
\label{av2}
\end{eqnarray}

The equality between Eqs \ref{av1}
and \ref{av2}
for any bra $\langle x_0 \vert $ and for any ket $\vert {\tilde x}_0 \rangle \! \rangle $
leads to the identity at the operator level
\begin{eqnarray}
 e^{t {\bold w}^{\dagger} }  {\bold \Omega} ={\bold \Omega} e^{t  {\tilde{\bold w}} }
\label{av12t}
\end{eqnarray}
that should be moreover valid for any time $t$, so that one obtains the following identity 
\begin{eqnarray}
  {\bold w}^{\dagger}   {\bold \Omega} ={\bold \Omega}   {\tilde{\bold w}} 
\label{av12}
\end{eqnarray}
that only involves the duality matrix ${\bold \Omega} $ and the two Markov generators ${\bold w} $ and ${\tilde{\bold w}} $, while the arbitrary time $t$ and the two arbitrary initial conditions $(x_0,{\tilde x_0})$ have disappeared (see the reviews \cite{ReviewMohle,ReviewDuality,AlgebraicReview} where this identity is written in various forms depending on the notations).


\subsection{ Markov duality in terms of the spectral decompositions of the two generators ${\bold w} $ and ${\tilde{\bold w}} $ }

To understand the meaning of the operator identity of Eq. \ref{av12},
it is useful to consider the spectral properties of the two Markov generators \cite{DualityEigen}.

\subsubsection{ Spectral decomposition of the Markov generator ${\bold w} $
 in the bi-orthogonal basis of its left and right eigenvectors}

 The general case involving degenerate eigenvalues and Jordan blocks has been analyzed in \cite{DualityEigen}.
 Here to simplify the discussion, we will focus on the cases where the eigenvalues $(-E)$
 of the Markov generator ${\bold w} $ are discrete and non-degenerate,
 so that they can be thus used to label the corresponding right and left eigenvectors.
Then the spectral decomposition of Markov generator ${\bold w} $
involves a sum over its eigenvalues $(-E)$ that may be complex
\begin{eqnarray}
{\bold w} = - \sum_{E } E \vert r_{E} \rangle \langle l_{E} \vert
\label{spectral}
\end{eqnarray}
  while
 the corresponding right eigenvectors $\vert r_{E} \rangle $ 
and left eigenvectors $\langle l_{E} \vert $ satisfy the eigenvalues equations
\begin{eqnarray}
  - E \vert r_{E} \rangle && =  {\bold w} \vert r_{E} \rangle
  \nonumber \\
 - E \langle l_{E} \vert && =  \langle l_{E} \vert{\bold w}  
\label{spectralrl}
\end{eqnarray}
and form a bi-orthogonal basis with the orthonormalization and closure relations
\begin{eqnarray}
\delta_{E,e} && = \langle l_{E} \vert  r_{e} \rangle = \sum_x \langle l_{E} \vert x \rangle \langle x \vert r_{e} \rangle
= \sum_x  l_{E}^* (x) r_{e}(x)
\nonumber \\
{\bold 1} && = \sum_{E}  \vert r_{E} \rangle \langle l_{E} \vert 
\ \ \ \ \ \ \ \text { or equivalently } \ \ 
\delta_{x,y} =  \sum_{E} \langle x \vert r_{E} \rangle \langle l_{E} \vert y \rangle
= \sum_{E}  r_{E}(x)  l_{E}^*(y)
\label{orthorl}
\end{eqnarray}

Let us write in coordinates the eigenvalue Eqs \ref{spectralrl} 
for the right eigenvector $\langle x \vert r_{E} \rangle = r_E(x) $ 
and the left eigenvector $ \langle l_{E} \vert x\rangle = l_E^*(x)$
\begin{eqnarray}
  - E r_E(x)  && = \sum_y w(x,y) r_E(y)
  \nonumber \\
 - E l_E^*(y) && = \sum_x l_E^*(x)w(x,y) 
\label{eigenleftcoordinate}
\end{eqnarray}
 The complex-conjugates of these two equations 
 read using that the matrix elements $w(x,y) $ of the Markov matrix are real
\begin{eqnarray}
  - E^* r_E^*(x)  && = \sum_y w(x,y) r_E^*(y)
  \nonumber \\
 - E^* l_E(y) && = \sum_x l_E(x)w(x,y) 
\label{eigenleftcoordinatecc}
\end{eqnarray}
So when $E$ is a complex eigenvalue, the complex-conjugate $E^*$ is also an eigenvalue
with the corresponding right and left eigenvectors
\begin{eqnarray}
  r_{E^*}(x) &&= r_E^*(x)  
  \nonumber \\
 l_{E^*}(y) && = l_E^*(y) 
\label{rlforEstar}
\end{eqnarray}

The spectral decomposition of Eq. \ref{spectral} is useful to rewrite the
propagator of Eq. \ref{propagator} as
\begin{eqnarray}
p_t(x \vert x_0) \equiv \langle x \vert e^{ t {\bold w} } \vert x_0 \rangle 
&& = \sum_{E} e^{-t E}  \langle x \vert r_{E} \rangle \langle l_{E} \vert x_0 \rangle
= \sum_{E} e^{-t E}  r_{E} (x)  l_{E}^*( x_0 )
\label{propagatorspectral}
\end{eqnarray}

From the point of view of the first process, the duality function $\Omega(x,{\tilde x}) =\langle x \vert {\bold \Omega} \vert {\tilde x} \rangle \! \rangle $ 
of Eq. \ref{dualityfunction} can be considered 
\begin{eqnarray}
\Omega(x,{\tilde x}) = \langle x \vert {\bold \Omega} \vert {\tilde x} \rangle \! \rangle 
\equiv \langle x \vert O_{[{\tilde x} ]} \rangle = O_{[{\tilde x} ]}(x)
\label{dualityfunction1}
\end{eqnarray}
as a collection of observables $O_{[{\tilde x} ]}(x)=\langle x \vert O_{[{\tilde x} ]} \rangle$ labelled by the parameter ${[{\tilde x} ]} $.

The averaged value ${\mathbb E}_{x_0} [ O(X(t)] $ 
  of an arbitrary observable $O(x)= \langle x \vert O \rangle$ 
  reads using the spectral decomposition of Eq. \ref{propagatorspectral}
\begin{eqnarray}
{\mathbb E}_{x_0} [ O(x(t)]  && =  \sum_x O(x) p_t(x \vert x_0)  
= \sum_x  \langle x \vert O \rangle
\sum_{E} e^{-t E}  \langle x \vert r_{E} \rangle \langle l_{E} \vert x_0 \rangle
\nonumber \\
&& = \sum_{E} e^{-t E}  \bigg( \sum_x  \langle x \vert O \rangle  \langle x \vert r_{E} \rangle
\bigg)  \langle l_{E} \vert x_0 \rangle
\label{observable1spectral}
\end{eqnarray}
and thus its dynamics a priori involves all the eigenvalues $E$.

However, for the observable $O(x)=l_{e}^*(x)$ corresponding to $\langle x \vert O \rangle =\langle l_{e} \vert x \rangle  $,
the dynamics of Eq. \ref{observable1spectral} 
\begin{eqnarray}
{\mathbb E}_{x_0} [ l_{e}^*(X(t)]  && 
=  \sum_{E} e^{-t E}  \bigg( \sum_x  \langle l_{e} \vert x \rangle  \langle x \vert r_{E} \rangle
\bigg)  \langle l_{E} \vert x_0 \rangle
=   \sum_{E} e^{-t E}  \  \langle l_{e} \vert r_{E} \rangle  \langle l_{E} \vert x_0 \rangle
\nonumber \\
&& =\sum_{E} e^{-t E}  \  \delta_{e,E}
   l_{E}^*(x_0 ) = e^{-t e}l_{e}^*(x_0 ) 
\label{observable1spectralleft}
\end{eqnarray}
reduces to the single exponential associated to the eigenvalues $e$.
 As a consequence, the left eigenvectors can be interpreted as a basis of simple observables
 whose dynamics involves a single exponential.


\subsubsection{ Spectral decomposition of the Markov generator ${\tilde {\bold w}} $
 in the bi-orthogonal basis of its left and right eigenvectors}

Similarly, one can write the spectral decomposition analog to Eq. \ref{spectral}
for the Markov generator ${\tilde w}$ of the dual process
\begin{eqnarray}
{\tilde {\bold w}} = - \sum_{\tilde E } {\tilde E} \vert {\tilde r}_{\tilde E} \rangle  \! \rangle
\langle  \! \langle {\tilde l}_{\tilde E} \vert
\label{spectraltilde}
\end{eqnarray}
in terms of its eigenvalues ${\tilde E} $ and in terms of the corresponding right eigenvectors $\vert {\tilde r}_{\tilde E} \rangle  \! \rangle $
and left eigenvectors $\langle  \! \langle {\tilde l}_{\tilde E} \vert $ 
that form a bi-orthogonal basis of the dual configuration space.

Then the propagator of Eq. \ref{propagatortilde} becomes
\begin{eqnarray}
{\tilde p}_t({\tilde x} \vert  {\tilde x}_0) \equiv 
 \langle  \! \langle {\tilde x} \vert e^{t {\tilde{\bold w}}} \vert {\tilde x}_0 \rangle \! \rangle
 = \sum_{\tilde E } e^{-t  {\tilde E} } \langle  \! \langle {\tilde x}
 \vert {\tilde r}_{\tilde E} \rangle  \! \rangle
\langle  \! \langle {\tilde l}_{\tilde E} \vert  {\tilde x}_0)
\rangle  \! \rangle
=\sum_{\tilde E } e^{-t  {\tilde E} }  {\tilde r}_{\tilde E} ({\tilde x})
 {\tilde l}_{\tilde E}^*(   {\tilde x}_0)
\label{propagatortildespectral}
\end{eqnarray}

From the point of view of the second process, the duality function $\Omega(x,{\tilde x}) =\langle x \vert {\bold \Omega} \vert {\tilde x} \rangle \! \rangle $ 
of Eq. \ref{dualityfunction} can be considered 
\begin{eqnarray}
\Omega(x,{\tilde x}) = \langle x \vert {\bold \Omega} \vert {\tilde x} \rangle \! \rangle 
\equiv \langle \! \langle {\tilde O}_{[x ]} \vert {\tilde x} \rangle \! \rangle= {\tilde O}_{[x ]}^*({\tilde x})
\label{dualityfunction2}
\end{eqnarray}
as a collection of observables ${\tilde O}_{[x ]}^*({\tilde x})=\langle \! \langle {\tilde O}_{[x ]} \vert {\tilde x} \rangle \! \rangle $ labelled by the parameter $x $.

The averaged value ${\mathbb E}_{{\tilde x_0}} [ O^*({\tilde X}(t)] $ 
  of an arbitrary observable ${\tilde O}^*({\tilde x})=\langle \! \langle {\tilde O} \vert {\tilde x} \rangle \! \rangle$ 
  reads using the spectral decomposition of Eq. \ref{propagatortildespectral}
  \begin{eqnarray}
{\mathbb E}_{{\tilde x_0}} [ O^*({\tilde X}(t)]  
&& = \sum_{\tilde x} O^*({\tilde x}) {\tilde p}_t({\tilde x} \vert  {\tilde x}_0)
= \sum_{\tilde E } e^{-t  {\tilde E} } 
\bigg( \sum_{\tilde x} \langle \! \langle {\tilde O} \vert {\tilde x} \rangle \! \rangle
 \langle  \! \langle {\tilde x} \vert {\tilde r}_{\tilde E} \rangle  \! \rangle \bigg)
\langle  \! \langle {\tilde l}_{\tilde E} \vert  {\tilde x}_0)
\rangle  \! \rangle
\nonumber \\
&& =  \sum_{\tilde E } e^{-t  {\tilde E} } 
 \langle \! \langle {\tilde O} \vert  {\tilde r}_{\tilde E} \rangle  \! \rangle 
\langle  \! \langle {\tilde l}_{\tilde E} \vert  {\tilde x}_0)
\rangle  \! \rangle
\label{observable2spectral}
\end{eqnarray}
and thus its dynamics a priori involves all the eigenvalues $\tilde E$.

However, for the observable $\langle \! \langle {\tilde O} \vert {\tilde x} \rangle \! \rangle=\langle \! \langle {\tilde l}_{\tilde e} \vert {\tilde x} \rangle \! \rangle$ corresponding to $ {\tilde O}^*(\tilde x) = {\tilde l}_{\tilde e}^*(\tilde x)  $,
the dynamics of Eq. \ref{observable2spectral} 
  \begin{eqnarray}
{\mathbb E}_{{\tilde x_0}} [ {\tilde l}_{\tilde e}^*({\tilde X}(t)]  
&&  =  \sum_{\tilde E } e^{-t  {\tilde E} } 
 \langle \! \langle {\tilde l}_{\tilde e} \vert  {\tilde r}_{\tilde E} \rangle  \! \rangle 
\langle  \! \langle {\tilde l}_{\tilde E} \vert  {\tilde x}_0)
\rangle  \! \rangle
=   \sum_{\tilde E } e^{-t  {\tilde E} } 
\delta_{\tilde E, \tilde e}
\langle  \! \langle {\tilde l}_{\tilde E} \vert  {\tilde x}_0)
\rangle  \! \rangle
\nonumber \\
&& = e^{-t  {\tilde e} }  {\tilde l}_{\tilde e}^* ( {\tilde x}_0)
\label{observable2spectralleft}
\end{eqnarray}
reduces to a single exponential as in Eq. \ref{observable1spectralleft}.
 Here again, the left eigenvectors can be interpreted as a basis of simple observables
for the dual process.


\subsubsection{ Spectral characterization of the possible dual matrices ${\bold \Omega}$ }

Plugging the spectral decompositions of Eqs \ref{spectral} and \ref{spectraltilde} for the two Markov generators ${\bold w} $ and ${\tilde {\bold w} } $
into the operator identity of Eq. \ref{av12} yields
\begin{eqnarray}
0 && =  {\bold w}^{\dagger}   {\bold \Omega} -{\bold \Omega}   {\tilde{\bold w}} 
\nonumber \\
&& = {\bold \Omega}  \sum_{\tilde E } {\tilde E} \vert {\tilde r}_{\tilde E} \rangle  \! \rangle
\langle  \! \langle {\tilde l}_{\tilde E} \vert
 - \sum_{E} E^* \vert l_{E} \rangle \langle r_{E} \vert  {\bold \Omega}
\label{av12spectral}
\end{eqnarray}

It is thus convenient to expand the duality matrix ${\bold \Omega} $ in the following 
basis of the left eigenvectors of the two generators
\begin{eqnarray}
 {\bold \Omega}   = \sum_{E} \sum_{{\tilde E}}
 \omega_{E,{\tilde E}}  \vert l_{E} \rangle \langle \! \langle {\tilde l}_{\tilde E} \vert 
\label{OmegaLL}
\end{eqnarray}
where the coefficients $\omega_{E,{\tilde E}} $ correspond to the following matrix elements
involving the right eigenvectors 
of the two generators
\begin{eqnarray}
 \omega_{E,{\tilde E}} = \langle r_{E} \vert {\bold \Omega} \vert {\tilde r}_{\tilde E} \rangle \! \rangle
\label{OmegaRR}
\end{eqnarray}

Then the projection of Eq. \ref{av12spectral}
on the bra $ \langle r_{e} \vert $ and on the ket $\vert {\tilde r}_{\tilde e} \rangle \! \rangle$ 
\begin{eqnarray}
0 && =\sum_{\tilde E } {\tilde E} 
\langle r_{e} \vert {\bold \Omega}  \vert {\tilde r}_{\tilde E} \rangle  \! \rangle
\langle  \! \langle {\tilde l}_{\tilde E} \vert {\tilde r}_{\tilde e} \rangle \! \rangle
 - \sum_{E} E^* 
\langle r_{e} \vert  l_{E} \rangle \langle r_{E} \vert  {\bold \Omega}\vert {\tilde r}_{\tilde e} \rangle \! \rangle
\nonumber \\
&&=\sum_{\tilde E } {\tilde E} 
\langle r_{e} \vert {\bold \Omega}  \vert {\tilde r}_{\tilde E} \rangle  \! \rangle
\delta_{\tilde E,\tilde e} 
 - \sum_{E} E^* \delta_{e,E}
\langle r_{E} \vert  {\bold \Omega}\vert {\tilde r}_{\tilde e} \rangle \! \rangle
\nonumber \\
&& = 
 {\tilde e} 
\langle r_{e} \vert {\bold \Omega}  \vert {\tilde r}_{\tilde e} \rangle  \! \rangle
 - e^*  \langle r_{e} \vert  {\bold \Omega}\vert {\tilde r}_{\tilde e} \rangle \! \rangle
 \nonumber \\
&& = (  {\tilde e}  -  e^*) \omega_{e,{\tilde e}}
\label{av12spectralRR}
\end{eqnarray}
yields that the coefficient $\omega_{e,{\tilde e}} $ can be non-vanishing
only if the corresponding eigenvalues are complex-conjugate ${\tilde e}  =  e^* $ to each other.

In conclusion, the decomposition of Eq. \ref{OmegaLL} for the duality matrix reduces to a single sum over
the eigenvalue $E={\tilde E}^*$
\begin{eqnarray}
 {\bold \Omega}   = \sum_{E } \omega_{E,E^*}  \vert l_{E} \rangle \langle \! \langle {\tilde l}_{E^*} \vert \ \ \ 
 \text{ with coefficients } \ \ \omega_{E,E^*} = \langle r_{E} \vert {\bold \Omega} \vert {\tilde r}_{E^*} \rangle \! \rangle
\label{OmegaLLpairing}
\end{eqnarray}

The physical interpretation is that 
the duality function $\Omega(x,{\tilde x}) =\langle x \vert {\bold \Omega} \vert {\tilde x} \rangle \! \rangle $ 
of Eq. \ref{dualityfunction} 
\begin{eqnarray}
\Omega(x,{\tilde x}) && = \langle x \vert {\bold \Omega} \vert {\tilde x} \rangle \! \rangle 
=  \sum_{E } \omega_{E,E^*}\langle x  \vert l_{E} \rangle \langle \! \langle {\tilde l}_{E^*} \vert  {\tilde x} \rangle \! \rangle
\nonumber \\
&& \equiv   \sum_{E } \omega_{E,E^*}O_{E,E^*}( x,{\tilde x}) 
\label{dualityfunctionCL}
\end{eqnarray}
can be considered as a linear combination with coefficients $\omega_{E,E^*} $ of the elementary 
observables $O_{E,E^*}( x,{\tilde x}) =  \langle x  \vert l_{E} \rangle \langle \! \langle {\tilde l}_{E^*} \vert  {\tilde x} \rangle \! \rangle$ that reduce to products of the left eigenvectors
that can be also rewritten in two other forms using the property $ l_{E^*}(y)  = l_E^*(y)  $ of Eq. \ref{rlforEstar}
with respect to complex-conjugation of the eigenvalues and eigenvectors
\begin{eqnarray}
 O_{E,E^*}( x,{\tilde x}) && \equiv  \langle x  \vert l_{E} \rangle \langle \! \langle {\tilde l}_{E^*} \vert  {\tilde x} \rangle \! \rangle
 =  l_{E} (x) {\tilde l}_{E^*}^*( {\tilde x} )
 =l_{E} (x) {\tilde l}_{E}( {\tilde x} )
 =  l_{E^*}^* (x) {\tilde l}_{E^*}^*( {\tilde x} )
 \label{Oleftleft}
\end{eqnarray}

The dynamics of these left eigenvectors obtained from Eqs \ref{observable1spectralleft}
and \ref{observable2spectralleft}
\begin{eqnarray}
{\mathbb E}_{x_0} [ l_{E}(X(t)]  && 
 = e^{-t E^*}l_{E}(x_0 ) 
 \nonumber \\
{\mathbb E}_{{\tilde x_0}} [ {\tilde l}_{E}({\tilde X}(t)]  
&& = e^{-t E^* }  {\tilde l}_{E} ( {\tilde x}_0)
\label{dynobservableleftleft}
\end{eqnarray}
involves the same single exponential time dependence $e^{-t E^*} $.

However, if one uses a single elementary function $O_{E,E^*}( x,{\tilde x})  $ of Eq. \ref{Oleftleft}
as duality function, the corresponding Markov duality simply means that the two models have a common eigenvalue 
and is thus not very informative.
As a consequence, the most interesting Markov dualities are between models 
that are isospectral, i.e. 
that have all their eigenvalues in common as described in the next subsection.


\subsection{ Duality functions between isospectral Markov models
involving only non-vanishing coefficients $\omega_{E,E^*} \ne 0 $ }

When the duality matrix $ {\bold \Omega}  $ is characterized by non-vanishing coefficients $ \omega_{E,E^*} \ne 0$ in Eq. \ref{dualityfunctionCL}, it is then useful to introduce the inverse
\begin{eqnarray}
 {\bold \Omega}^{-1}   = \sum_{e } \frac{1}{ \omega_{e,e^*} } \vert {\tilde r}_{e^*} \rangle \! \rangle  \langle r_{e} \vert 
\label{OREALinverse}
\end{eqnarray}
satisfying
\begin{eqnarray}
{\bold \Omega} {\bold \Omega}^{-1} &&  = 
 \sum_{E }  \sum_{e } \frac{ \omega_{E,E^*}}{ \omega_{e,e^*} }
  \vert l_{E} \rangle \langle \! \langle {\tilde l}_{E^*} \vert 
 \vert {\tilde r}_{e^*} \rangle \! \rangle  \langle r_{e} \vert 
 =  \sum_{E }  
  \vert l_{E} \rangle   \langle r_{E} \vert 
 ={\bold 1}
 \nonumber \\
 {\bold \Omega}^{-1} {\bold \Omega} &&  = 
  \sum_{e }  \sum_{E } \frac{ \omega_{E,E^*}}{ \omega_{e,e^*} }
 \vert {\tilde r}_{e^*} \rangle \! \rangle  \langle r_{e} \vert 
  \vert l_{E} \rangle \langle \! \langle {\tilde l}_{E^*} \vert 
  =   \sum_{e } 
 \vert {\tilde r}_{e^*} \rangle \! \rangle   \langle \! \langle {\tilde l}_{e^*} \vert 
   = \tilde {{\bold 1}}
\label{OREALinversecheck}
\end{eqnarray}
Then the operator identity of Eq. \ref{av12} can be multiplied by the inverse ${\bold \Omega}^{-1}  $ 
on the right or on the left to obtain the similarity transformations
\begin{eqnarray}
  {\bold w}^{\dagger} &&  ={\bold \Omega}   {\tilde{\bold w}} {\bold \Omega}^{-1}
  \nonumber \\
 {\tilde{\bold w}}  && =  {\bold \Omega}^{-1}  {\bold w}^{\dagger}   {\bold \Omega}   
\label{av12inverse}
\end{eqnarray}


\subsection{ Choice of normalization for the eigenvectors to have all coefficients equal to unity $\omega_{E,E^*} =1 $ }

The coefficients $ \omega_{E,E^*} = \langle r_{E} \vert {\bold \Omega} \vert {\tilde r}_{E^*} \rangle \! \rangle$ of Eq. \ref{OmegaLLpairing}
are related to the choice of normalization for the right eigenvectors of the two models,
while the bi-orthonormalization of Eq. \ref{orthorl}
only determines the relations between the right and the left eigenvectors
of each model.
As a consequence, it is possible to choose the normalization of eigenvectors of the two models
to have all coefficients equal to unity  
 \begin{eqnarray}
 \omega_{E,E^*} =1
\label{coefsunity}
\end{eqnarray}
so that the duality function of Eq. \ref{OmegaLLpairing}
\begin{eqnarray}
 {\bold \Omega}   = \sum_{E }   \vert l_{E} \rangle \langle \! \langle {\tilde l}_{E^*} \vert 
\label{OmegaLLbijective}
\end{eqnarray}
can be then considered as a mapping between the right and the left eigenvectors of the two models
in the following sense
\begin{eqnarray}
\langle r_e \vert  {\bold \Omega} = \sum_{E } \langle r_e  \vert l_{E} \rangle \langle \! \langle {\tilde l}_{E^*} \vert 
= \langle \! \langle {\tilde l}_{e^*} \vert 
\nonumber \\
 {\bold \Omega} \vert {\tilde r}_{e^*} \rangle \! \rangle  
 = \sum_{E }   \vert l_{E} \rangle \langle \! \langle {\tilde l}_{E^*} \vert {\tilde r}_{e^*} \rangle \! \rangle
 =\vert l_{e} \rangle
\label{OmegaLLbijectiveaction}
\end{eqnarray}
while the inverse of Eq. \ref{OREALinverse} that reduces to
\begin{eqnarray}
 {\bold \Omega}^{-1}   = \sum_{e }  \vert {\tilde r}_{e^*} \rangle \! \rangle  \langle r_{e} \vert 
\label{OREALinversebijective}
\end{eqnarray}
is useful to invert the relations of Eqs \ref{OmegaLLbijectiveaction}
\begin{eqnarray}
\langle r_e \vert 
= \langle \! \langle {\tilde l}_{e^*} \vert  {\bold \Omega}^{-1}
\nonumber \\
 \vert {\tilde r}_{e^*} \rangle \! \rangle  
 =  {\bold \Omega}^{-1}\vert l_{e} \rangle
\label{OmegaLLbijectiveactionInverse}
\end{eqnarray}

Let us write the relations of Eqs \ref{OmegaLLbijectiveaction}
in coordinates using Eq. \ref{rlforEstar}
\begin{eqnarray}
{\tilde l}_e({\tilde x}) &&= {\tilde l}_{e^*}^*({\tilde x})  = \sum_x r_e^*(x)   \Omega (x,{\tilde x})    
\nonumber \\
l_{e} (x) && =  \sum_{{\tilde x}}  \Omega(x,{\tilde x})  {\tilde r}_{e^*}({\tilde x})  
 = \sum_{{\tilde x}}  \Omega(x,{\tilde x})  {\tilde r}_{e}^*({\tilde x}) 
\label{OmegaLLbijectiveactionCoordinates}
\end{eqnarray}
since they will be useful for the various examples discussed in the next sections.
In many of these examples, the eigenvalues and the eigenvectors of the two Markov generators are real, 
so that one will be able to drop the 
complex-conjugate notations that are needed when complex eigenvalues are present.


\section{ Spectral interpretation of known Markov dualities in the same configuration space}

\label{sec_TimeSiegmund}

In this section,  two well-known Markov dualities between processes defined in the same configuration space
are revisited via the spectral perspective in order to characterize the simple links between the right and left eigenvectors
of the two generators.

\subsection{ Time-Reversal-Duality as a exchange between right and left eigenvectors via the steady state $p_{ss}(x)$}

Let us consider the case where the Markov chain of Eq. \ref{master} converges towards a steady state $p_{ss}(x)$
satisfying
\begin{eqnarray}
 0  =\partial_t p_{ss}(x) = \sum_{y } w(x,y) p_{ss}( y )
\label{mastereq}
\end{eqnarray}
while Eq. \ref{wdiag} can be rewritten as
\begin{eqnarray}
0  =  \sum_{x} w(x,y) 
\label{wdiag0}
\end{eqnarray}
These two equations mean that $E=0$ is an eigenvalue for the Markov matrix ${\bold w}$
with the corresponding right and left eigenvectors
\begin{eqnarray}
   r_{E=0} (x)&& =p_{ss}(x)
  \nonumber \\
 l_{E=0} (x)&& =1
\label{rlzero}
\end{eqnarray}
The spectral decomposition of Eq. \ref{propagatorspectral} then describes the relaxation towards the steady state
$p_{ss}(x)  $ for any intial condition $x_0$
\begin{eqnarray}
p_t(x \vert x_0)  
&& = \sum_{E} e^{-t E}  \langle x \vert r_{E} \rangle \langle l_{E} \vert x_0 \rangle
\nonumber \\
&& = p_{ss}(x) +  \sum_{E \ne 0} e^{-t E}  \langle x \vert r_{E} \rangle \langle l_{E} \vert x_0 \rangle
\label{propagatorverssteady}
\end{eqnarray}

The Markov matrix ${\tilde w}$ associated to the time-reversed dynamics 
\begin{eqnarray}
{\tilde w}(y,x) \equiv \frac{ w(x,y) p_{ss}(y) }{ p_{ss}(x)}   
 \label{Rreversed}
\end{eqnarray}
has also $E=0$ as eigenvalue with the left eigenvector ${\tilde l}_0(x)=1 $
and the right eigenvector ${\tilde r}_0(x)=p_{ss}(x) $ as can be checked
\begin{eqnarray}
\sum_{y} {\tilde w}(y,x) && =  \frac{ 1 }{ p_{ss}(x)} \sum_{y} w(x,y) p_{ss}(y) =0
\nonumber \\
\sum_{x} {\tilde w}(y,x) p_{ss}(x) && = p_{ss}(y)  \sum_x w(x,y)    = 0
 \label{RMarkov}
\end{eqnarray}
The propagator $ {\tilde p}_t(x_0 \vert x ) $ of the reversed dynamics of generator ${\tilde w}$ is directly
related to the propagator of $p_t(x \vert x_0) $  of Eq. \ref{propagatorverssteady} via
\begin{eqnarray}
  {\tilde p}_t(x_0 \vert x )  =    \frac{ p_t(x \vert x_0) p_{ss}(x_0) }{ p_{ss}(x)}
 && =  \sum_{E} e^{-t E} \bigg( \frac{ \langle x \vert r_{E} \rangle }{p_{ss}(x)} \bigg) \bigg( \langle l_{E} \vert x_0 \rangle p_{ss}(x_0) \bigg)
 \nonumber \\
 && \equiv \sum_{E} e^{-t  E} \langle x_0 \vert {\tilde r_{E}} \rangle  \langle {\tilde l_{E}  } \vert x \rangle 
 \label{Rreversedpropagator}
\end{eqnarray}
As a consequence, the eigenvalues $E$ are the same in the two models, 
while the corresponding right and left eigenvectors are related via
\begin{eqnarray}
\langle x_0 \vert {\tilde r_{E}} \rangle  && =   \langle l_{E}  \vert x_0 \rangle p_{ss}(x_0) 
\ \ \ \ \text{ i.e. } \ \ {\tilde r_{E}}(x_0) = l_E^*(x_0) p_{ss}(x_0)
  \nonumber \\
 \langle {\tilde l_{E}  } \vert x \rangle && =  \frac{ \langle x \vert r_{E} \rangle}{p_{ss}(x)} 
 \ \ \ \ \ \ \ \ \ \ \ \ \text{ i.e. } \ \ \tilde l_{E}^*(x) = \frac{ r_E(x)}{p_{ss}(x) }
 \label{eigenreversed}
\end{eqnarray}

The comparison with Eqs \ref{OmegaLLbijectiveactionCoordinates}
\begin{eqnarray}
 {\tilde l}_e(y) && = \sum_x r_e^*(x)   \Omega (x,y)   
\nonumber \\
 l_{e} (x) && = \sum_{y}  \Omega(x,y)  {\tilde r}_{e}^*(y)  
\label{OmegaLLbijectiveactionCoordinatesTimeReversal}
\end{eqnarray}
leads to the well-known diagonal duality function that involves the steady state $p_{ss}(x) $
\begin{eqnarray}
 \Omega (x,y) = \frac{\delta_{x,y}}{p_{ss}(x)}
\label{OmegaTimeReversal}
\end{eqnarray}


\subsection{ Siegmund-Duality where left eigenvectors correspond to integrals of the dual right eigenvectors }

As recalled in the Introduction with many references, the Siegmund-Duality
is one of the oldest and most studied duality.
Since the Siegmund duality has already been analyzed from the spectral point of view
for Markov jump processes in \cite{DualityEigen},
let us focus here instead on  
the Siegmund duality between two diffusion processes defined on the interval $x \in [0,L]$,
where the eigenvalues and the eigenvectors of the two generators are all real.

The Siegmund duality function is the Heaviside function (whose derivative is the delta function)
\begin{eqnarray}
 \Omega_S( x,y) = \theta(x-y )
 \label{DSiegmund}
\end{eqnarray}
As a consequence, the translation of Eqs \ref{OmegaLLbijectiveactionCoordinates}
for the continuous interval $[0,L]$ yields 
that the left real eigenvectors correspond to integrals of the right real eigenvectors
\begin{eqnarray}
 {\tilde l}_e(y) &&= \int_0^L dx \ r_e(x)   \Omega_S (x,y)   = \int_y^L dx \ r_e(x) 
\nonumber \\
 l_{e} (x) &&= \int_0^L dy \  \Omega_S(x,y)  {\tilde r}_{e}(y)  = \int_0^x dy \  {\tilde r}_{e}(y)
\label{OmegaLLbijectiveactionCoordinatesSiegmund}
\end{eqnarray}
while the right eigenvectors correspond to derivatives of the left eigenvectors
\begin{eqnarray}
\frac{d {\tilde l}_e(y)}{dy} &&  =- r_e(y) 
\nonumber \\
\frac{d l_{e} (x) }{dx} &&  =   {\tilde r}_{e}(x)
\label{OmegaLLbijectiveactionCoordinatesSiegmundderi}
\end{eqnarray}

The eigenvalue equation for the left eigenvector $l_E(x)$ that involves the adjoint operator of Eq. \ref{FokkerPlanckAdjoint}
\begin{eqnarray}
-E l_E(x) = {\cal F}^{\dagger} l_E(x) = F_I(x) \frac{ d  l_E(x)}{dx} + D(x)  \frac{ d^2  l_E(x)}{dx^2}
\label{Lefteigensieg}
\end{eqnarray}
can be derived to obtain the following equation for the dual right eigenvector
$ {\tilde r}_{E} (x) =  \frac{d l_E(  x ) }{d x}    $ of Eq. \ref{OmegaLLbijectiveactionCoordinatesSiegmundderi}
\begin{eqnarray}
-E {\tilde r}_{E} (x) && = \frac{d}{dx} \left[  F_I(x) {\tilde r}_{E} (x) + D(x)  \frac{ d {\tilde r}_{E} (x)}{dx} \right]
\nonumber \\
&& = \frac{d}{dx}  \left[\bigg(F_I(x) -D'(x) \bigg) {\tilde r}_{E} (x) \right]+   \frac{ d^2 }{dx^2} \bigg(D(x)  {\tilde r}_{E} (x)\bigg)
\label{Lefteigensiegtilde}
\end{eqnarray}

The comparison with the eigenvalue Equation for the right eigenvector $ {\tilde r}_{E} (x)$
that involves the dual Fokker-Planck generator ${\tilde {\cal F} }$ of Eq. \ref{FokkerPlanck}
containing the dual Ito force ${\tilde F}_I(x)$ and the dual diffusion coefficient ${\tilde D}(x) $
\begin{eqnarray}
-E {\tilde r}_{E} (x)  &&={\tilde {\cal F} } {\tilde r}_{E} (x)
 = - \frac{d}{dx} \bigg( {\tilde F}_I(x) {\tilde r}_{E} (x) \bigg) + 
 \frac{ d^2 }{dx^2} \bigg( {\tilde D}(x) {\tilde r}_{E} (x) \bigg) 
\label{FokkerPlanckdual}
\end{eqnarray}
yields that the diffusion coefficients coincide ${\tilde D}(x) =D(x) $ ,
while the Ito forces of the two models are related via 
\begin{eqnarray}
 {\tilde F}_I(x) = - F_I(x) + D'(x) 
\label{FItoDual}
\end{eqnarray}
Note that in terms of the Fokker-Planck forces $F(x)=F_I(x)-D'(x)$ and ${\tilde F}(x)={\tilde F}_I(x)-D'(x) $
of the two models used in \cite{c_boundarydriven,c_susyFokkerPlanck},
the duality transformation reads instead 
\begin{eqnarray}
{\tilde F}(x)={\tilde F}_I(x)-D'(x) =  - F_I(x) = -F(x)-D'(x)
\label{FFPDual}
\end{eqnarray}


\subsection{ Discussion }

In the two examples of the present section, the Markov duality was 
between two Markov processes living in the same configuration space,
and the spectral analysis was useful to write very simple relations 
between the right and left eigenvectors in the two models.
In the next section, we turn to the case where the two Markov processes have 
 different configuration spaces,
namely a continuous interval and a discrete lattice.


\section{ Moments-Duality $\Omega(x,n)  = x^n$ between diffusion processes in $x$ 
and Markov jump processes in $n \in {\mathbb N}$ }

\label{sec_moment}

In the present section, we focus on the moments duality
based on the duality function (see the reviews \cite{ReviewMohle,ReviewDuality,AlgebraicReview} and references therein)
\begin{eqnarray}
  \Omega (x,n)  =x^n
\label{OmegaM}
\end{eqnarray}
between diffusion processes in the continuous space $x$ governed by a Fokker-Planck generator ${\cal F} $ 
and a Markov jump process governed by a Markov matrix ${\tilde w} (m,n)$ on the semi-infinite lattice $n \in {\mathbb N}$, so that the operator identity of Eq. \ref{av12} has to be adapted into
\begin{eqnarray}
  {\cal F}^{\dagger}  \Omega(x,n) = \sum_{m}  \Omega (x,m)  {\tilde w} (m,n)
\label{av12diffusion}
\end{eqnarray}
We first focus on the well-known duality concerning the Wright-Fisher diffusion
before discussing possible generalizations.


\subsection{ Moments-Duality $\Omega(x,n)  = x^n$ between the Wright-Fisher diffusion on the interval $x \in [0,1] $ and the Kingman Markov jump process on the semi-infinite lattice $n \in {\mathbb N}$ }

\label{subsec_WF}

The Moments-Dualities have been much studied in the context of population genetics \cite{ReviewMohle,Redig_genetic}
and in particular for the Wright-Fisher diffusion (see the review \cite{ReviewWF} and references therein).


\subsubsection{ Wright-Fisher diffusion on $x \in [0,1] $ with the diffusion coefficient $D(x)=x(1-x) $
and the Ito Force $F_I(x)=\alpha (1-x)$ }

The Wright-Fisher diffusion with mutation on the interval $x \in [0,1] $ involves the quadratic diffusion coefficient $D(x)=x(1-x) $
and the Ito Force $F_I(x)=\alpha (1-x)$ where $\alpha>0$ is the positive mutation parameter. 
So the adjoint operator ${\cal F}^{\dagger}$ of Eq. \ref{FokkerPlanckAdjoint}
\begin{eqnarray}
 {\cal F}^{\dagger} =  F_I(x) \frac{\partial}{\partial x } +D(x) \frac{\partial^2}{\partial x^2 }  
 = \alpha (1-x) \frac{\partial}{\partial x } +  x(1-x) \frac{\partial^2}{\partial x^2 } 
\label{WFgene}
\end{eqnarray}
can be applied to the duality function $\Omega(x,n)  = x^n $ 
to obtain the left handside of Eq. \ref{av12diffusion}
\begin{eqnarray}
 {\cal F}^{\dagger} \Omega(x,n)
&& = \alpha (1-x) \frac{\partial}{\partial x } x^n + x(1-x) \frac{\partial^2}{\partial x^2 } x^n
=\alpha (1-x) n x^{n-1} + x(1-x) n (n-1) x^{n-2} 
\nonumber \\
&& = \bigg[ \alpha n +n (n-1)  \bigg] (x^{n-1} - x^n) \equiv \lambda_n (x^{n-1} - x^n)
\label{WFgeneO}
\end{eqnarray}
where we have introduced the notation 
\begin{eqnarray}
\lambda_n \equiv \alpha n +n (n-1)  = n  \bigg( \alpha  + (n-1)  \bigg) >0 \ \ \text{ for $n>0$ when $\alpha>0$}
\label{WFlambdan}
\end{eqnarray}
The 'no-mutation' case $\alpha=0$ can also be considered but leads to $\lambda_{n=1}=0$,
so to simplify the present discussion we will focus on the case $\alpha>0$ where the only vanishing parameter is
$\lambda_{n=0}=0$.

Let us label the real eigenvalues $(-E_k)$ by the integer $k \in {\mathbb N}$ and write the 
eigenvalues equations for the corresponding real left and right eigenvectors
\begin{eqnarray}
- E_k l_k(x) &&= {\cal F}^{\dagger} l_k(x) =  \alpha (1-x) \frac{\partial l_k(x) }{\partial x } + x(1-x) \frac{\partial^2  l_k(x)}{\partial x^2 } 
\nonumber \\
- E_k r_k(x) &&= {\cal F} r_k(x) = - \frac{\partial}{\partial x } \bigg(  \alpha (1-x) r_k(x) \bigg) + 
  \frac{\partial^2 }{\partial x^2 } \left[ x(1-x) r_k(x) \right]
\label{WFeigen}
\end{eqnarray}

The vanishing energy $E_{k=0}=0$ is associated to the absorbing state at the boundary $x=1$
\begin{eqnarray}
r_0(x) && =\delta(x-1)
\nonumber \\
l_0(x) && =1
\label{WFAbsorbing}
\end{eqnarray}


\subsubsection{ The Kingman Markov jump process on the semi-infinite lattice $n \in {\mathbb N}$ }

The Kingman Markov jump process on the semi-infinite lattice $n \in {\mathbb N}$
 is defined by the Markov matrix that involves the parameters $\lambda_n$ of Eq. \ref{WFlambdan}
\begin{eqnarray}
{\tilde w}(m,n) = \lambda_n \bigg( \delta_{m,n-1}- \delta_{m,n}\bigg)
\label{WKingman}
\end{eqnarray}
This directed model analyzed in detail in Appendix \ref{app_Directed} can be understood as follows:
when the particle is on the site $n \in {\mathbb N}$, the only possible move is a jump towards 
the left neighbor $m=n-1$ with the positive transition rate ${\tilde w}(n-1,n) =\lambda_n>0 $,
while $\lambda_0=0$ means that $n=0$ is an absorbing state.
The right handside of Eq. \ref{av12diffusion}
\begin{eqnarray}
 \sum_{m}  \Omega (x,m)  {\tilde w} (m,n) = \sum_m x^m  \ \lambda_n \bigg( \delta_{m,n-1}- \delta_{m,n}\bigg)
 =  \lambda_n (x^{n-1} - x^n)
\label{av12diffusionw}
\end{eqnarray}
indeed coincides with Eq. \ref{WFgeneO}, so that the operator identity of Eq. \ref{av12diffusion} is satisfied.

As described in Appendix \ref{app_Directed} for the case of arbitrary distinct transition rates $\lambda_n >0$ for $n>0$ while $\lambda_{n=0}=0$,
the eigenvalues $(-E_k)$ of Eq. \ref{EkMarkov}
are simply given by the transition rates $\lambda_k$
\begin{eqnarray}
E_k= \lambda_k
\label{KingEklambdak}
\end{eqnarray}
while  the corresponding right eigenvectors $ {\tilde r}_{k} (n)  $ are written in Eq. \ref{EigenRprod}
and the left eigenvectors $  {\tilde l}_{k} (n)$ are written in Eq. \ref{EigenLprod}.


\subsubsection{ Relations between the right and left eigenvectors of the two isospectral models}

The relations of Eq. \ref{OmegaLLbijectiveactionCoordinates}
read for the present Markov models involving real eigenvectors 
\begin{eqnarray}
 {\tilde l}_k(n) && =\int_0^1 dx \ r_k(x)   \Omega (x,n) =\int_0^1 dx \ r_k(x)  x^n
\nonumber \\
 l_k (x) && = \sum_{n=0}^{+\infty}  \Omega(x,n) \  {\tilde r}_k(n)  =  \sum_{n=0}^{+\infty}  x^n \  {\tilde r}_k(n)
\label{OmegaLLbijectiveactionCoordinatesMoments}
\end{eqnarray}

As a consequence, the explicit expressions for the eigenvectors ${\tilde r}_k$
 and ${\tilde l}_k $ of the dual model given in Appendix \ref{app_Directed}
can be used to analyze the eigenvectors $r_k(x) $ and $l_k(x) $ of the Wright-Fisher diffusion process as follows:
 
 (i) Plugging the dual right eigenvector ${\tilde r}_k(n)$
 of Eq. \ref{EigenRprod} into Eq. \ref{OmegaLLbijectiveactionCoordinatesMoments}
yields that  the left eigenvector $ l_k (x) $ of the Wright-Fisher diffusion 
 \begin{eqnarray}
 l_k (x) &&   = x^k \  {\tilde r}_k(k) + \sum_{n=0}^{k-1}  x^n \  {\tilde r}_k(n) 
 \nonumber \\
 && = x^k  + \sum_{n=0}^{k-1}  x^n \left( \prod_{j=n}^{j=k-1}  \frac{ \lambda_{j+1}   }   {(\lambda_{j} - \lambda_{k} ) } \right)
\label{lkdiffusion}
\end{eqnarray}
reduces to a polynomial of degree $k$ with explicit coefficients.

 (ii) Plugging the dual left eigenvector ${\tilde l}_k(n)$
 of Eq. \ref{EigenLprod} into Eq. \ref{OmegaLLbijectiveactionCoordinatesMoments}
yields that  the right eigenvector $ r_k (x) $ of the Wright-Fisher diffusion
can be characterized by the following explicit integrals for $n \in {\mathbb N}$
\begin{eqnarray}
\int_0^1 dx \ r_k(x)  x^n && =  {\tilde l}_k(n)= 0 \ \ \ \ \text { for $0 \leq n<k$ }
\nonumber \\
\int_0^1 dx \ r_k(x)  x^k && =  {\tilde l}_k(k)= 1 
\nonumber \\
\int_0^1 dx \ r_k(x)  x^n && =  {\tilde l}_k(n) = \prod_{j=k+1}^n    \frac{ \lambda_{j}   }   {(\lambda_{j} - \lambda_{k} ) }  
   \ \  \ \ \text{ for } 
     n \geq k+1
\label{rkdiffusion}
\end{eqnarray}


\subsection{ Generalization of the moments duality to other Pearson diffusions with quadratic $D(x)$ and linear $F_I(x)$}

Since the Wright-Fisher diffusion belongs to the family of Pearson diffusions 
with quadratic $D(x)$ and linear $F_I(x)$
that have many common remarkable properties that explain their exact-solvability
\cite{pearson1895,pearson_wong,diaconis,autocorrelation,pearson_class,pearson2012,PearsonHeavyTailed,pearson2018,c_pearson}, a natural question is whether the previous moments duality analysis can be generalized to other Pearson diffusions.

\subsubsection{ Generalization of the moments duality to other Jacobi diffusions with $D(x)=x(1-x)$ on the interval $x \in [0,1]$}

The Wright-Fisher diffusion belongs to the family of Jacobi diffusions on the interval $x \in [0,1]$
that involves the same diffusion coefficient $D(x)=x(1-x)$,
while the Ito force $F_I(x)$ involves the additional parameter $\beta$ 
\begin{eqnarray}
    F_I(x)  =  \alpha - (\alpha+\beta) x
\label{forcesjacobi}
\end{eqnarray}
 with respect to the case $\beta=0$ of the Wright-Fisher diffusion considered above.

Note that when the two parameters $(\alpha,\beta)$ are strictly positive, 
the diffusion converges towards the normalizable equilibrium steady state
\begin{eqnarray}
r_0(x) = p_{eq}(x)=
\frac{\Gamma(\alpha+\beta)}{ \Gamma(\alpha) \Gamma(\beta) } x^{\alpha-1} (1-x)^{\beta-1}
 \ \ \ {\rm for } \ x \in ]0,1[
 \ \ \ {\rm if } \ \ \alpha>0 \ \ {\rm and } \ \ \beta>0
\label{jacobi}
\end{eqnarray}
instead of the absorbing state of Eq. \ref{WFAbsorbing}.

In the presence of the parameter $\beta$, Eq. \ref{WFgeneO} becomes
\begin{eqnarray}
 {\cal F}^{\dagger} \Omega(x,n)
&& = \left[  \alpha - (\alpha+\beta) x \right] \frac{\partial}{\partial x } x^n + x(1-x) \frac{\partial^2}{\partial x^2 } x^n
= \left[  \alpha - (\alpha+\beta) x \right] n x^{n-1} + x(1-x) n (n-1) x^{n-2} 
\nonumber \\
&& = \bigg( \alpha n +n (n-1)  \bigg) x^{n-1} 
- \bigg( \alpha n +n (n-1) + \beta n \bigg) x^n
\nonumber \\
&& \equiv \lambda_n x^{n-1} - \left( \lambda_n + \beta n \right) x^n
\label{MomentJacobi}
\end{eqnarray}
in terms of the paramterers $\lambda_n=n  \left[ \alpha  + (n-1)  \right] >0 $
of Eq. \ref{WFlambdan}.
As a consequence, the dual generator of Eq. \ref{WKingman} has to be modified into
\begin{eqnarray}
{\tilde w}(m,n) = \lambda_n \bigg( \delta_{m,n-1}- \delta_{m,n}\bigg) - \beta n \delta_{m,n}
\label{WKingmanJacobi}
\end{eqnarray}
in order to obtain that the equivalence with Eq. \ref{MomentJacobi}
\begin{eqnarray}
 \sum_{m}  \Omega (x,m)  {\tilde w} (m,n) = 
 \sum_m x^m  \  \left[ \lambda_n \delta_{m,n-1}- \left( \lambda_n + \beta n \right)\delta_{m,n}\right]
 =  \lambda_n x^{n-1} - \left( \lambda_n + \beta n \right) x^n
\label{av12diffusionJacobi}
\end{eqnarray}
So in addition to the directed transition rates ${\tilde w}(n-1,n) = \lambda_n$
discussed in the previous subsection and in the Appendix, 
the dual generator contains the killing rates $K_n \equiv \beta n$ 
on the sites $n \in {\mathbb N}^* $
for $\beta >0$ (while for $\beta<0$, $K_n$ should be interpreted as reproducing rate $(-\beta n)>0$).

This example shows that when one tries to generalize a known Markov duality, it is very easy to 
obtain a dual generator ${\tilde w} $ that does not conserve the probability, i.e. the diagonal matrix elements
do not satisfy Eq. \ref{wdiag} but contain additional contributions that can be interpreted as killing rates or reproducing rates.


\subsubsection{ Generalization of the moments duality to Pearson diffusions with $D(x)=ax^2+bx$ on the interval $x \in [0,+\infty[$}

Let us now consider the Pearson diffusions on the infinite interval $x \in [0,+\infty[$,
where the Ito force is linear
\begin{eqnarray}
    F_I(x)  =  \alpha - \gamma x
\label{forcespearson}
\end{eqnarray}
while the quadratic positive diffusion coefficient $D(x)$ vanishes at $x=0$
\begin{eqnarray}
    D(x)  = a x^2 + b x \ \ \ \text { with $a \geq 0$ and $b \geq 0$ while $(a,b) \ne (0,0)$ }
\label{Dpearson}
\end{eqnarray}

The analog of Eq. \ref{WFgeneO} reads
\begin{eqnarray}
 {\cal F}^{\dagger} \Omega(x,n) && = \left[  F_I(x) \frac{\partial}{\partial x } +D(x) \frac{\partial^2}{\partial x^2 }\right] x^n 
= (\alpha - \gamma x) n x^{n-1} + (a x^2 + b x) n (n-1) x^{n-2} 
\nonumber \\
&& = \bigg( \alpha n +b n (n-1)  \bigg) x^{n-1} - \bigg( \gamma n -a n (n-1)  \bigg) x^n 
\label{MomentsPearson}
\end{eqnarray}

As a consequence, in order to reproduce this result via
\begin{eqnarray}
 \sum_{m}  \Omega (x,m)  {\tilde w} (m,n) = 
 \sum_m x^m  \  {\tilde w} (m,n)
 = {\tilde w} (n-1,n) x^{n-1} + {\tilde w} (n,n) x^n
\label{av12pearson}
\end{eqnarray}
the dual generator ${\tilde {\bold w}} $ should have the off-diagonal elements generalizing 
the transition rates of Eq. \ref{WFlambdan}
\begin{eqnarray}
\lambda_n ={\tilde w}(n-1,n)  = n \bigg( \alpha  +b  (n-1)  \bigg) >0
\text{ for $n>0$ when $\alpha>0$ and $b\geq 0$}
\label{offdiagpearson}
\end{eqnarray}
while the diagonal elements generalizing Eq. \ref{av12diffusionw}
\begin{eqnarray}
{\tilde w}(n,n) = - n \bigg( \gamma  -a  (n-1)  \bigg) \equiv - \lambda_n - K_n
\label{diagpearson}
\end{eqnarray}
do not a priori satisfy Eq. \ref{wdiag}
but involve the killing rates $K_n$ (or reproducing rates $(-K_n)>0$ if $K_n<0$)
\begin{eqnarray}
K_n && = - {\tilde w}(n,n)   - \lambda_n = n \bigg( \gamma  -a  (n-1)  \bigg) - n \bigg( \alpha  +b  (n-1)  \bigg)
\nonumber \\
&& =  n \bigg( (\gamma- \alpha)  - (a +b)  (n-1)  \bigg) 
\label{killingpearson}
\end{eqnarray}
Note that the coefficient $(-a-b)$ of the quadratic term $n^2$ can never vanish 
for the case of the positive diffusion coefficient  $D(x)  = a x^2 + b x $ on $x \in ]0,+\infty[$ of Eq. \ref{Dpearson},
in contrast to the Wright-Fisher diffusion corresponding to $b=1=-a$.


\subsubsection{ Discussion}

In summary, the moments duality based on the duality function $\Omega(x,n)=x^n$ 
between the Wright-Fisher diffusion with mutation and the Directed Markov jump process
 on the semi-infinite lattice $n \in {\mathbb N}$
cannot be directly generalized to other Pearson diffusions,
unless one is ready to introduce additional killing rates or reproducing rates in the dual Markov jump model
as explained above.
As a consequence, it is natural to ask whether one should instead change the duality function $\Omega(x,n) $
in order to define Markov dualities between Pearson diffusions and appropriate Directed jump models that
conserve the probability. 
In the next section, we thus consider the more general problem of 
constructing new Markov dualities from the spectral properties of the two models. 


\section{ Constructing an appropriate dual Directed Jump Model in Eigenspace  }

\label{sec_DirectedEigen}

The goal of this section is to analyze whether the spectral perspective 
can be used to construct new dualities and to give a simple example.

 \subsection{ Discussion : using the spectral perspective to construct new dualities} 

In the two previous sections, well-known Markov dualities based on simple duality functions 
 $\Omega(x,\tilde x)$ defined in the configuration spaces $(x,\tilde x )$ of the two models 
have been revisited from the spectral point of view.
In the present section, the goal is instead to use the spectral perspective in order to construct new dualities.
To be concrete, we will focus on the following question :
if a Markov process is given with its generator ${\bold w}$ and 
spectral decomposition of Eq. \ref{spectral}
that involves its eigenvalues $(-E)$ and its corresponding right and left eigenvectors,
what is the simplest dual model that can be constructed ?

The dual generator ${\tilde {\bold w}}$ should have exactly the same eigenvalues $(-E)$,
but one has some freedom to choose the dual configuration space ${\tilde x}$ and the biorthogonal basis 
of the left eigenvectors ${\tilde l}_{E} ({\tilde x})$ and the right eigenvectors ${\tilde r}_{E}({\tilde x})$
in order to define the dual generator ${\tilde {\bold w}}$ via its spectral decomposition of Eq. \ref{spectraltilde}
\begin{eqnarray}
{\tilde {\bold w}} = - \sum_{ E } E \vert {\tilde r}_E \rangle  \! \rangle
\langle  \! \langle {\tilde l}_E \vert
\label{spectraltildedef}
\end{eqnarray}
However, the non-trivial constraints are that any off-diagonal element ${\tilde x} \ne {\tilde y}$
of the Markov generator ${\tilde {\bold w}} $ in its configuration space should be positive
\begin{eqnarray}
\langle  \! \langle  {\tilde x} \vert {\tilde {\bold w}} \vert {\tilde y} \rangle \! \rangle
= - \sum_{ E }E \langle  \! \langle  {\tilde x}
\vert {\tilde r}_E \rangle  \! \rangle
\langle  \! \langle {\tilde l}_E \vert
{\tilde y} \rangle \! \rangle
\geq 0 
\ \ \ \ \text{for any 
${\tilde x} \ne {\tilde y}$ }
\label{so}
\end{eqnarray}
in order to represent a genuine transition rate from ${\tilde y} $ to ${\tilde x} $,
while the conservation of probability can be taken 
into account by imposing the left eigenvector ${\tilde l}_{E=0}({\tilde x})=1$ associated to $E=0$.

In the next subsection, we thus focus on the case 
where the dual process ${\tilde {\bold w}}  $ is a Directed Markov jump process on $n \in {\mathbb N}$
as already encountered in the previous section, but here the only hypothesis is that 
the generator ${\bold w} $ has non-degenerate real eigenvalues,
instead of being the specific Wright-Fisher diffusion considered in the previous section.


\subsection{ Case of non-degenerate real eigenvalues : Markov duality towards a Directed Jump model in Eigenspace}

Let us consider that the Markov process takes place in a space of $N$ configurations,
and that the Markov matrix ${\bold w}$ of size $N \times N$ has $N$ distinct real eigenvalues $(-E_k) \leq 0$
labelled by the integer $k \in \{0,1,..,N-1\}$ with
\begin{eqnarray}
E_0=0<E_1<E_2<..<E_{N-1}
\label{Eclassement}
\end{eqnarray}
where the vanishing eigenvalue $E_{k=0}=0$ is associated to some steady state $p_{ss}(x)$
\begin{eqnarray}
r_0(x) && =p_{ss}(x)
\nonumber \\
l_0(x) && =1
\label{r0l0ss}
\end{eqnarray}
On can also consider the case $N=+\infty$ when the discrete configuration space becomes infinite,
or when the Markov matrix ${\bold w} $ is replaced by some Fokker-Planck generator in continuous space.

One can use the analysis of the Directed Markov jump processes of Appendix \ref{app_Directed},
in particular the eigenvalues of Eq. \ref{EkMarkov} as follows.
One can consider the Directed Markov jump process
where the transition rates $\lambda_k$ are chosen to be given by the energies $E_k$ of Eq. \ref{Eclassement}
so that the generator of Eq. \ref{W} reads
\begin{eqnarray}
 {\tilde w } (q,k)   =   E_k (\delta_{q,k-1} -  \delta_{q,k})
\label{Weigen}
\end{eqnarray}
with the following interpretation :

(i)  The configuration space labelled by the integer $k \in\{0,1,..,N-1\}$ represents the Eigenspace.

(ii) The dual generator ${\tilde {\bold w} } $ of Eq. \ref{Weigen} is characterized by the transition rates $ {\tilde w } (k-1,k) =E_k$
that are also the opposite-eigenvalues of the dual model as explained in Eq. \ref{EkMarkov}.

(iii) The dual process corresponds to a particle that visits the various excited energies 
in the order $(E_{N-1} \to E_{N-2} \to ... \to E_2 \to E_1 \to E_0=0$ of Eq. \ref{Eclassement}
 and that spends
 at level $E_k$ a random time $t_k \in [0,+\infty[$ drawn with the exponential distribution 
\begin{eqnarray}
 P_k(t_k)   =   E_k e^{- t E_k}
\label{PktExponential}
\end{eqnarray}
before jumping to the level $E_{k-1}$, and this process continues up to the final absorbing state $E_0=0$.

(iv) When starting at the highest levels $E_{N-1}$, 
the Mean-First-Passage-Time ${\tilde \tau}_{Abs}$ at the absorbing state $E_0=0$ 
obtained from Eq. \ref{PktExponential}
reduces to the sum of the inverse of the non-vanishing energies $E_k >0$
\begin{eqnarray}
 {\tilde \tau}_{Abs} = \sum_{k=1}^{N-1} \frac{1}{E_k} = \tau_{Kemeny}
\label{Kemeny}
\end{eqnarray}
and thus coincides with the Kemeny-time $\tau_{Kemeny} $ that characterizes the convergence towards the steady state $p_{ss}(x)$ of the first Markov process (see detailed discussions with examples in \cite{us_kemeny} and references therein).

(v) The duality function $ \Omega(x,q)  $ is given in terms of the left eigenvectors of the two models via Eq. \ref{OmegaLLbijective}
\begin{eqnarray}
 \Omega(x,q)   = \sum_{k=0}^{N-1}   l_k(x)  {\tilde l}_k(q) 
\label{OmegaLLbijectiveDirected}
\end{eqnarray}
where the left eigenvectors ${\tilde l}_k(q) $ of the dual model of Eq. \ref{EigenLprod} read for the present case with
$\lambda_k = E_k$
\begin{eqnarray}
 {\tilde l}_{k}(q) && = 0  \ \ \ \ \text{ for $0 \leq q< k$ }
\nonumber \\
{\tilde l}_{k}(k) && =1
\nonumber \\
    {\tilde l}_{k} (q) &&  = \prod_{j=k+1}^q    \frac{ E_{j}   }   {(E_{j} - E_{k} ) }  
    \ \ \text{ for } 
      k+1 \leq q \leq N-1
 \label{EigenLprodDir}
\end{eqnarray}
As a consequence, the duality function of Eq. \ref{OmegaLLbijectiveDirected}
contains only the eigenvectors $l_k(x)$ labelled by $0 \leq k \leq q$
\begin{eqnarray}
 \Omega(x,q)   =  l_q(x)    + \sum_{k=0}^{q-1}   l_k(x)   \prod_{j=k+1}^q    \frac{ E_{j}   }   {(E_{j} - E_{k} ) }  
\label{OmegaLLbijectiveDirectedFinite}
\end{eqnarray}

When the left eigenvectors $ l_k(x) $ associated to the generator ${\bold w}$ are not known,
the duality function $\Omega(x,q)  $ of Eq. \ref{OmegaLLbijectiveDirected}
remains somewhat implicit in terms of the coordinates  $(x,q)$ of the two configuration spaces.

For the examples of Pearson diffusions where the left eigenvector $l_k(x) $ is a known polynomial of degree $k$
\cite{pearson1895,pearson_wong,diaconis,autocorrelation,pearson_class,pearson2012,PearsonHeavyTailed,pearson2018,c_pearson},
one obtains that the duality function $ \Omega(x,q)  $ of Eq. \ref{OmegaLLbijectiveDirectedFinite}
is a polynomial of degree $q$,
that generalizes the special case $  \Omega(x,q)=x^q$ of the Wright-Fisher diffusion
discussed in the previous section.

In the last subsection \ref{subsec_meta} of the Appendix, we describe the simplifications when 
the energies $E_n$ are ordered 
and very separated $E_{n+1} \gg E_n$,
i.e. when the Markov dynamics of the two models involves a series of well-separated metastable states.


\section{ Conclusions}

\label{sec_conclusion}

In this paper, we have revisited the notion of Markov duality via the spectral decompositions of the two Markov generators in their bi-orthogonal basis of right and left eigenvectors. While the general case involving degenerate eigenvalues and Jordan blocks has been analyzed in the mathematical paper \cite{DualityEigen}, 
we have focused here
 on the cases where the eigenvalues $(-E)$ of the Markov generators are non-degenerate in order to obtain a simple picture : the two generators should have the same eigenvalues $(-E)$ that may be complex, while the duality function $\Omega(x,{\tilde x})$ can be considered as a mapping between the right and the left eigenvectors of the two models. 
 This general framework has been first illustrated on the well-known examples of
 the Time-Reversal duality and of the Siegmund duality, where the two configurations spaces are the same
 and where the relations between the left and right eigenvectors of the two models are very simple.
 We have then focused on the Moment-Duality between the Wright-Fisher diffusion on the interval $x \in [0,1] $ and the Kingman Markov jump process on the semi-infinite lattice $n \in {\mathbb N}$ in order to analyze the relations between their eigenvectors living in two different configuration spaces. Finally, we have discussed how the spectral perspective can be used to construct new dualities and we have given an example for the case of non-degenerate real eigenvalues, where one can always construct a dual Directed Jump process on the semi-infinite lattice $n \in {\mathbb N}$, whose transitions rates are the opposite-eigenvalues.
We thus hope that other new Markov dualities will be constructed in the future via this spectral perspective.


\appendix

\section{ Directed Markov jump process on the half-line $n \in {\mathbb N}$ towards the absorbing state $n=0$} 

\label{app_Directed}

In this Appendix, we describe the Directed Markov jump process on the half-line $n \in {\mathbb N}$  
with the transition rates $\lambda_n$ 
\begin{eqnarray}
{\tilde w}(n-1,n) =\lambda_n>0  \ \ \ \text{for $n>0$}
\label{deflambdan}
\end{eqnarray}
towards the absorbing state $n=0$ characterized by
\begin{eqnarray}
{\tilde w}(-1,0) =\lambda_0 =0
\label{deflambdan0}
\end{eqnarray}
Since this model will be used as the dual process in the two sections \ref{sec_moment}
and \ref{sec_DirectedEigen} of the main text with different interpretations and different parameters,
we will assume in this Appendix that the transition rates $\lambda_n$ are distinct but otherwise arbitrary.

Note that in the vocabulary of birth-death models, the above model is a pure-death model,
while in the field of slow glassy dynamics, where trap models have been much studied in various versions
\cite{jpb_weak,dean,jp_ac,jp_pheno,bertinjp1,bertinjp2,trapsymmetric,trapnonlinear,trapreponse,trap_traj,c_ruelle,c_ring},
the above model corresponds to a Directed Trap model.


\subsection{ Markov matrix ${\tilde w }(.,.)$ for the Directed Markov jump process   
with arbitrary transition rates $\lambda_n$ }

The matrix elements of the Markov matrix 
\begin{eqnarray}
 {\tilde w } (m,n)   =   \lambda_n (\delta_{m,n-1} -  \delta_{m,n})
\label{W}
\end{eqnarray}
lead to the master equation of Eq. \ref{master} for the propagator ${\tilde p_t }(n \vert n_0)$ 
\begin{eqnarray}
\partial_t {\tilde p_t }(n \vert n_0)  = \sum_{n'} {\tilde w}(n,n') {\tilde p_t }(n' \vert n_0) 
 = \lambda_{n+1} {\tilde p_t }(n+1\vert n_0)-  \lambda_{n} {\tilde p_t }(n\vert n_0) 
\label{MasterDeath2}
\end{eqnarray}
where the first term for $n=0$ involving $\lambda_{n=0}=0$ 
\begin{eqnarray}
\partial_t {\tilde p_t }(0\vert n_0) && = \lambda_1 {\tilde p_t }(1\vert n_0)
\label{MasterDeath012}
\end{eqnarray}
describes the irreversible absorption at $n=0$.
So the right and left eigenvectors associated to the vanishing eigenvalue $E=0$ read
\begin{eqnarray}
 {\tilde r}_0(n) && =\delta_{n,0}
 \nonumber \\
  {\tilde l}_0(n_0) && =1
\label{r0l0}
\end{eqnarray}

\subsection{ Spectral decomposition governing the convergence towards the absorbing state }

In the main text, we have chosen to write the spectral decompositions of Eq. \ref{spectral}
as a sum over the eigenvalues $(-E)$ in order to stress the difference with sums involving the
configuration space.
However here, it is more convenient to label the real eigenvalues $(-E_k)$ via the integer $k \in {\mathbb N}$
with the corresponding right eigenvectors $  {\tilde r}_k$ and left eigenvectors $ {\tilde l}_k $
satisfying the eigenvalues Equations of Eq. \ref{spectralrl}
\begin{eqnarray}
 -E_{k} \vert {\tilde r}_{k} \rangle \! \rangle &&= {\tilde w} \vert {\tilde r}_{k} \rangle \! \rangle
 \nonumber \\
  -E_{k} \langle \! \langle {\tilde l}_{k} \vert && =\langle \! \langle {\tilde l}_{k} \vert  {\tilde w}  
 \label{keigen}
\end{eqnarray}
and the bi-orthonormalization and closure relations of Eq. \ref{orthorl}
\begin{eqnarray}
\delta_{k, k' } && = \langle \! \langle {\tilde l}_{k} \vert {\tilde r}_{k'} \rangle \! \rangle
= \sum_{n=0}^{+\infty}{\tilde l}_{k}(n)  {\tilde r}_{k'}(n)
\nonumber \\
\delta_{n,m} && =\sum_{k=0}^{+\infty} \langle \!  \langle n \vert {\tilde r}_k \rangle \! \rangle
\langle \! \langle {\tilde l}_k \vert m \rangle \! \rangle
=  \sum_{k=0}^{+\infty}    {\tilde r}_{k}(n)  {\tilde l}_{k}(m)
 \label{orthoLRbeta}
\end{eqnarray}

The spectral decomposition of Eq. \ref{propagatortildespectral}
for the propagator 
\begin{eqnarray}
{\tilde p_t }(n \vert n_0) 
 = \sum_{k=0}^{+\infty} e^{- t E_{k}}  {\tilde r}_{k}(n) {\tilde l}_{k}(n_0)
 = \delta_{n,0} + \sum_{k=1}^{+\infty} e^{- t E_{k}}  {\tilde r}_{k}(n) {\tilde l}_{k}(n_0)
\label{DirectedPropagator}
\end{eqnarray}
 describes the convergence towards the absorbing state at $n=0$ for any initial condition $n_0$.


\subsection{ Explicit eigenvalues and supports of the right and the left eigenvectors}

The rewriting of the eigenvalues Eqs \ref{keigen}
for the real eigenvectors in coordinates
\begin{eqnarray}
 -E_{k}  {\tilde r}_{k} (n) &&=   \sum_{n'}  {\tilde w}(n,n') {\tilde r}_{k} (n')
 =  \lambda_{n+1} {\tilde r}_{k} (n+1) -   \lambda_{n}  {\tilde r}_{k} (n) 
 \nonumber \\
  -E_{k}  {\tilde l}_{k} (n) && =  \sum_m {\tilde l}_{k} (m) {\tilde w}   (m,n)
  =   \lambda_n {\tilde l}_{k} (n-1) - \lambda_n {\tilde l}_{k} (n)
 \label{EigenWRL}
\end{eqnarray}
show the triangular structure of both systems, so that the eigenvalues $(-E_k)$
are simply given by the diagonal terms $(-\lambda_k)$
\begin{eqnarray}
 E_{k} = \lambda_{k}  
 \label{EkMarkov}
\end{eqnarray}

In addition, the directed nature of the process shows that the propagator of Eq. \ref{DirectedPropagator}
can only involve the eigenvalues $(-E_k)$ with $k \in \{n,n+1,..,n_0-1,n_0\}$
\begin{eqnarray}
{\tilde p_t }(n \vert n_0) 
 = \sum_{k=n}^{n_0} e^{- t E_{k}}  {\tilde r}_{k}(n) {\tilde l}_{k}(n_0)
\label{DirectedPropagatorFinite}
\end{eqnarray}
Moreover for the given eigenvalue $(-E_k)$,
the right eigenvector 
\begin{eqnarray}
\text{ ${\tilde r}_{k}(n)$ has support $ n \in \{0,1,..,k \}$ and vanishes $ {\tilde r}_{k}(n) = 0 $ for $n> k$}
\label{Rbetazero}
\end{eqnarray}
while the left eigenvector
\begin{eqnarray}
\text{${\tilde l}_{k}(n_0) $ has support support $ n_0 \in \{k,k+1,k+2, ... \}$ 
and vanishes $ {\tilde l}_{k}(n_0) = 0 $ for $0 \leq n_0< k$ }
\label{Lbetazero}
\end{eqnarray}
As a consequence, 
the bi-orthonormalization of Eq. \ref{orthoLRbeta} for $k'=k$
 reduces to the single term $n=k$
\begin{eqnarray}
1= \langle \! \langle {\tilde l}_{k} \vert {\tilde r}_{k} \rangle \! \rangle
= \sum_{n=0}^{+\infty}{\tilde l}_{k}(n)  {\tilde r}_{k}(n)
= {\tilde l}_{k}(k) {\tilde r}_{k}(k)
 \label{orthoLRbetanorma}
\end{eqnarray}
and it is thus convenient to choose the following normalization for the eigenvectors
\begin{eqnarray}
 {\tilde l}_{k}(k) =1={\tilde r}_{k}(k)
 \label{orthoLRbetachoice}
\end{eqnarray}


\subsection{ Explicit computation of the right eigenvectors ${\tilde r}_{k}(n)$ }

With the property of Eq. \ref{Rbetazero},
the eigenvalue Eq. \ref{EigenWR} for the right eigenvector ${\tilde r}_{k}(.)$
\begin{eqnarray}
(\lambda_{n} -E_{k} ) {\tilde r}_{k} (n) && =  \lambda_{n+1} {\tilde r}_{k} (n+1) 
 \label{EigenWR}
\end{eqnarray}
 reduces to the finite system for $n=0,1,..,k$
\begin{eqnarray}
(\lambda_0 -E_{k} ) {\tilde r}_{k} (0) && =  \lambda_{1} {\tilde r}_{k} (1) 
 \nonumber \\
(\lambda_1 -E_{k} ) {\tilde r}_{k} (1) && =  \lambda_{2} {\tilde r}_{k} (2) 
  \nonumber \\
...
 \nonumber \\
(\lambda_{k-1} -E_{k} ) {\tilde r}_{k} (k-1) && =  \lambda_{k} {\tilde r}_{k} (k) 
\nonumber \\
(\lambda_{k} -E_{k} ) {\tilde r}_{k} (k) && =    0 
 \label{EigenWRfinite}
\end{eqnarray}
The last equation reproduces the eigenvalue $E_k=\lambda_k$ of Eq. \ref{EkMarkov},
while the other equations can be solved recursively backwards to obtain 
with the normalization choice of Eq. \ref{orthoLRbetachoice} 
\begin{eqnarray}
{\tilde r}_{k}(k) && =1
\nonumber \\
  {\tilde r}_{k} (k-1) && =     \frac{ \lambda_{k} }{\lambda_{k-1} - \lambda_{k}}  
\nonumber \\
 {\tilde r}_{k} (k-2) && =  
   \frac{ \lambda_{k-1}  \lambda_{k}  }
   {(\lambda_{k-2} - \lambda_{k} )( \lambda_{k-1} - \lambda_{k})  }  
\nonumber \\
 {\tilde r}_{k} (n) && =  
 \prod_{j=n}^{j=k-1}  \frac{ \lambda_{j+1}   }   {(\lambda_{j} - \lambda_{k} ) }  
 \ \ \ \text{ for } \ \ 0 \leq n \leq k-1
 \label{EigenRprod}
\end{eqnarray}


\subsection{ Explicit computation of the left eigenvectors ${\tilde l}_{k}(n)$ }

With the property of Eq. \ref{Lbetazero},
the eigenvalue Eq. \ref{EigenWL} for the left eigenvector ${\tilde l}_{k}(.)$
\begin{eqnarray}
(\lambda_n  -E_{k})  {\tilde l}_{k} (n)      =   \lambda_n {\tilde l}_{k} (n-1) 
 \label{EigenWL}
\end{eqnarray}
 becomes the following system for $n=k,k+1,k+2,...+\infty$
\begin{eqnarray}
 (\lambda_{k}  -E_{k})  {\tilde l}_{k} (k)   &&   =   0
       \nonumber \\
(\lambda_{k+1}  -E_{k})  {\tilde l}_{k} (k+1) &&     =   \lambda_{k+1} {\tilde l}_{k} (k) 
     \nonumber \\
(\lambda_{k+2}  -E_{k})  {\tilde l}_{k} (k+2)  &&    =   \lambda_{k+2} {\tilde l}_{k} (k+1) 
     \nonumber \\
     ...    
 \label{EigenWLfinite}
\end{eqnarray}
The first equation reproduces the eigenvalue $E_k=\lambda_k$ of Eq. \ref{EkMarkov},
while the other equations can be solved recursively to obtain 
with the normalization choice of Eq. \ref{orthoLRbetachoice} 
\begin{eqnarray}
{\tilde l}_{k}(k) && =1
\nonumber \\
  {\tilde l}_{k} (k+1) &&  =  \frac{   \lambda_{k+1}    }{ \lambda_{k+1} -\lambda_{k}}
\nonumber \\
  {\tilde l}_{k} (k+2) &&  =  \frac{ \lambda_{k+2}   \lambda_{k+1}    }
  {(\lambda_{k+2} -\lambda_{k}) (\lambda_{k+1} -\lambda_{k})}
\nonumber \\
    {\tilde l}_{k} (n) &&  = \prod_{j=k+1}^n    \frac{ \lambda_{j}   }   {(\lambda_{j} - \lambda_{k} ) }  
    \ \ \text{ for } 
     n \geq k+1
 \label{EigenLprod}
\end{eqnarray}



\subsection{ Simplifications when the transition rates $\lambda_n$ are ordered 
and very separated $\lambda_{n+1} \gg \lambda_n$ }

\label{subsec_meta}

All the previous results are valid for arbitrary transitions rates $\lambda_n$.
Let us now consider the special case where the transition rates $\lambda_n$ are ordered 
and very separated 
\begin{eqnarray}
 \lambda_0=0 \ll \lambda_1 \ll \lambda_2 \ll ... \ll \lambda_n \ll \lambda_{n+1} \ll ...
 \label{OrderLambda}
\end{eqnarray}

Then the right eigenvector $ {\tilde r}_{k} (n) $ of Eq. \ref{EigenRprod} 
\begin{eqnarray}
{\tilde r}_{k}(k) && =1
\nonumber \\
  {\tilde r}_{k} (k-1) && =     \frac{ 1 }{ \frac{\lambda_{k-1} } {\lambda_{k}} -1 }  
  = -1 + O \left( \frac{\lambda_{k-1} } {\lambda_{k}}\right)
\nonumber \\
 {\tilde r}_{k} (k-2) && =  
   \frac{  \frac{\lambda_{k-1} } {\lambda_{k}}    }
   {( \frac{\lambda_{k-2} } {\lambda_{k}} - 1 )( \frac{ \lambda_{k-1}}{ \lambda_{k} } -1) }  
   = O \left( \frac{\lambda_{k-1} } {\lambda_{k}}\right)
\nonumber \\
 {\tilde r}_{k} (n) && =    \frac{ 1 }{ \frac{\lambda_{k-1} } {\lambda_{k}} -1 } 
 \prod_{j=n}^{j=k-2}  \frac{ \frac{\lambda_{j+1}}{\lambda_k}   }   {( \frac{\lambda_{j}}{\lambda_k}  -1 ) }  
 = O \left( \prod_{j=n}^{j=k-2}   \frac{\lambda_{j+1}}{\lambda_k}   \right)
 \ \ \ \text{ for } \ \ 0 \leq n \leq k-1
 \label{EigenRprodMeta}
\end{eqnarray}
can be approximated at leading order where one 
keeps only the finite contributions
by
\begin{eqnarray}
{\tilde r}_{k}(n)  \simeq \delta_{n,k} - \delta_{n,k-1}
 \label{EigenRprodMetaApprox}
\end{eqnarray}

The left eigenvector $ {\tilde l}_{k} (n) $ of Eq. \ref{EigenLprod} 
\begin{eqnarray}
{\tilde l}_{k}(k) && =1
\nonumber \\
  {\tilde l}_{k} (k+1) &&  =  \frac{   1    }{ 1 - \frac{ \lambda_{k}}{\lambda_{k+1}} } 
  = 1 + O \left( \frac{\lambda_{k} } {\lambda_{k+1}}\right)
\nonumber \\
  {\tilde l}_{k} (k+2) &&  =  \frac{ 1      }
  {(1 - \frac{ \lambda_{k}}{\lambda_{k+2}}) (1 - \frac{ \lambda_{k}}{\lambda_{k+1}})}
  = 1 + O \left( \frac{\lambda_{k} } {\lambda_{k+1}}\right)+ O \left( \frac{\lambda_{k} } {\lambda_{k+2}}\right)
\nonumber \\
    {\tilde l}_{k} (n_0) &&  = \prod_{j=k+1}^{n_0}    \frac{ 1  }   {(1 - \frac{ \lambda_{k}}{\lambda_j} ) }  
    =  1 +\sum_{j=k+1}^{n_0} O \left( \frac{\lambda_{k} } {\lambda_{j}}\right)
    \ \ \text{ for } 
     n_0 \geq k+1
 \label{EigenLprodMeta}
\end{eqnarray}
can be approximated at leading order where one 
keeps only the finite contributions
by
\begin{eqnarray}
    {\tilde l}_{k} (n_0)   \simeq \theta(  n_0 \geq k)
 \label{EigenLprodMetaApprox}
\end{eqnarray}

Plugging the approximations of Eqs \ref{EigenRprodMetaApprox}
and \ref{EigenLprodMetaApprox}
into the spectral decomposition of Eq. \ref{propagatortildespectral}
for the propagator 
\begin{eqnarray}
{\tilde p_t }(n \vert n_0) && = \sum_{k=0}^{+\infty} e^{- t E_{k}}  {\tilde r}_{k}(n) {\tilde l}_{k}(n_0)
\simeq \sum_{k=0}^{+\infty} e^{- t E_{k}}  \left[ \delta_{n, k}  - \delta_{n,k-1} \right] \theta(  n_0 \geq k)
\nonumber \\
&& = e^{- t E_n}   \theta(  n_0 \geq n) - e^{- t E_{n-1}}   \theta(  n_0 \geq n-1)
\ \ \ \text{ with $E_{n-1}=\lambda_{n-1} \ll \lambda_n=E_n$ }
\label{Wspectralmeta}
\end{eqnarray}
corresponds to the decomposition of the dynamics in terms of well-separated metastable states.
The characterization of long-lived metastable states 
before the relaxation towards the Boltzmann equilibrium or some other steady state
has attracted a lot of interest for many classical stochastic dynamics
\cite{gaveau1,gaveauSchulman,gaveau2,GaveauMoreau,gaveau3,ReviewBiroli,bovier,bovier2,c_metastable,c_eigenvalue,c_RGdyn,c_Dyson,c_largestbarrier,c_LRSG}
(see also \cite{metastableQuantum_short,metastableQuantum_long} for 
the notion of metastability in open quantum systems).

For the analysis described in section \ref{sec_DirectedEigen}
of the main text,
the approximation of Eq. \ref{EigenLprodMetaApprox} for the dual left eigenvectors ${\tilde l}_{k} (.) $
 yields that the duality function of Eqs \ref{OmegaLLbijectiveDirected}
\ref{OmegaLLbijectiveDirectedFinite}
\begin{eqnarray}
 \Omega(x,q)   = \sum_{k=0}^{q}   l_k(x)  
\label{OmegaLLbijectiveDirectedMeta}
\end{eqnarray}
is simply given by the sum of the first $(q+1)$ left eigenvectors $ l_k(x)$.
 


\end{document}